\documentclass[reprint,aps,prl,amsmath,amssymb,superscriptaddress]{revtex4-2}

\usepackage{graphicx}
\usepackage{bm}
\usepackage{mathtools}
\usepackage{amsmath}
\usepackage{esvect}
\usepackage{placeins}
\usepackage{afterpage}
\usepackage{ragged2e}
\usepackage{color}
\usepackage[hidelinks]{hyperref}

\newcommand{\E}{\mathbb{E}}

\begin{document}

\title{A Utility-Driven Bounded-Confidence Model for Opinion Dynamics}

\author{Alex Siebenmorgen}
\email{alex.siebenmorgen@colorado.edu}
\affiliation{Department of Applied Mathematics, University of Colorado at Boulder, Boulder, CO 80309, USA}

\author{Juan G. Restrepo}
\email{juanga@colorado.edu}
\affiliation{Department of Applied Mathematics, University of Colorado at Boulder, Boulder, CO 80309, USA}

\date{\today}

\begin{abstract}
We introduce a utility-driven bounded-confidence model of opinion dynamics in which opinions associated with higher utility exert stronger social influence. In the regime where all agents belong to a single opinion cluster, we derive a stochastic differential equation for the mean opinion and show that its stationary distribution is Gibbs-like, with an effective potential determined by the utility landscape and an inverse temperature controlled by the learning rate and the number of agents. For multimodal utility functions, the dynamics exhibit metastability and spontaneous switching between competing opinion states. The reduced stochastic description also captures the evolution and merging of multiple opinion clusters, in agreement with agent-based simulations.
\end{abstract}

\maketitle

The evolution and dynamics of opinions in social networks underlie many important issues in modern society, such as the emergence of polarization and echo chambers \cite{bail2018exposure,baumann2020modeling,brooks2024emergence}, the effectiveness of misinformation campaigns \cite{bradshaw2018global}, and the response to public health measures in the face of disease outbreaks \cite{lang2021maskon,qiu2022understanding,cotfas2021longest}. Modeling the dynamics of opinions has, therefore, attracted the attention of researchers in many fields and is now a very active area of research \cite{sirbu2016opinion,peralta2025opinion,starnini2025opinion}. Opinion dynamics models usually represent an individual's opinion with a real-valued or binary value, and postulate mathematical rules for how an individual's opinion changes upon interaction with another. Some models with binary opinions include the voter model and its variations \cite{redner2019reality}, while models with continuous opinions include the deGroot consensus model \cite{french1956formal, degroot1974reaching}, the Friedkin–Johnsen
model \cite{friedkin1990social}, and bounded-confidence models \cite{deffuant2000mixing,rainer2002opinion, lorenz2007continuous}. In bounded-confidence models, a real-valued opinion is assigned to each individual, and two interacting individuals modify their opinions only if the difference between their opinions is small enough. These models assume that agents interact only with others whose opinions are sufficiently similar, and that interacting individuals partially compromise toward one another. 

Models of opinion dynamics do not typically make any assumption about which opinion might be preferable. In many settings, however, beliefs are not merely internal states: they are operationalized through actions (e.g., bets, decisions, predictions, expressions of belief) whose realized payoffs may provide feedback to the agent about the validity of their underlying opinion. In this Letter, we introduce a utility function that quantifies the relative desirability of opinion states. Our use of a utility function is motivated by the decision-theoretic tradition associated with Ramsey and Savage \cite{ramsey1931truth,savage2012foundations}, in which subjective probabilities are linked to preferences over actions and outcomes. In that spirit, we treat opinions as guiding actions that generate realized payoffs, and assume that opinions associated with higher utility exert stronger social influence. We modify the standard bounded confidence model opinion update rule so that the opinion with higher utility is more attractive. By analyzing a noisy version of this model, we show that opinion clusters evolve following a low-dimensional stochastic differential equation toward a Gibbs-like stationary density peaked around the local maxima of the utility function. Within the resulting small-spread approximation with additive independent noise, the stationary density depends only on the utility function, the number of agents, and the learning rate; the noise amplitude controls the cluster spread but cancels from the leading-order stationary density of the mean opinion. We demonstrate how this formalism can be used to simulate and study the evolution and merging of multiple opinion clusters, and to study spontaneous transitions of the mean population's opinion from one local maximum of the utility function to another.

\paragraph{Model.}
We consider a population of $N$ agents holding opinions that evolve in discrete time steps, $t = 0,\Delta t,2 \Delta t,...$, where $\Delta t=1/N$. We denote the opinion of agent $i$ at time $t$ by $x_i^t$. Although the opinions in this model can be interpreted as subjective probabilities or credences, for mathematical convenience we do not restrict them to the interval $[0,1]$. However, in our examples, they will satisfy $x_i^t \in [0,1]$ with very high probability. At each step, a pair $(i,j)$ is selected uniformly at random. If the pair satisfies the condition $|x_i^t-x_j^t|<\varepsilon$, where $\varepsilon>0$ is the \textit{confidence bound}, they interact; otherwise no update occurs. Upon interaction, agent $i$ updates its opinion via
\begin{equation}
x^{t+1}_{i}
=
x^t_i
+2\mu \frac{U_j({\mathbf{x}}^t)}{U_i({\mathbf{x}}^t)+U_j({\mathbf{x}}^t)}(x^t_j-x^t_i)
+\eta^t_i ,
\label{eq:update}
\end{equation}
and analogously for $j$. Here $\mu\in(0,1)$ is a learning rate and $\eta_i^t$ is a zero-mean independent noise term satisfying $\E[\eta_i^t]=0$ and $\E[\eta_i^t\eta_j^{t'}]=\Delta^2\delta_{ij} \delta_{tt'}$, where $\delta_{ij}$ is the Kronecker delta. The noise term represents effects that are not accounted for in the deterministic model, such as imperfect communication between agents or external influences.
The \textit{utility function} $U_i(\mathbf{x}^t)$ represents the strictly positive, cardinal payoff that agent $i$ obtains from acting on its current opinion $x_i^t$ given the opinions $\mathbf{x}^t=[x_1^t,x_2^t,...,x^t_N]^T$ held by the entire population. Traditional bounded-confidence models, in this framework, would implicitly assume that all opinions lead to the same utility [$U_i({\mathbf{x}}^t) \equiv 1$ for all $i$] \cite{deffuant2000mixing,rainer2002opinion}. In contrast, here we will consider the case where the utility depends directly on the agent's opinion, so that $U_i({\mathbf{x}}^t) \equiv U(x_i^t)$. Under the update rule in \eqref{eq:update}, the opinions of two agents $i$ and $j$ after an interaction satisfy 
\begin{equation}
x^{t+1}_{i}-x^{t+1}_j
= (1-2\mu)(x^t_i-x^t_j)+\eta^t_i-\eta^t_j.
\label{eq:avg_change}
\end{equation}
Therefore, since $|1-2\mu|<1$, their opinions move toward each other on average. However, the agent who has an opinion yielding a higher utility undergoes a lower magnitude revision. Thus the update rule is asymmetric, with the mean opinion of the two agents moving toward the opinion with the higher utility.

\begin{figure}[t]
  \centering
  \includegraphics[width=\columnwidth]{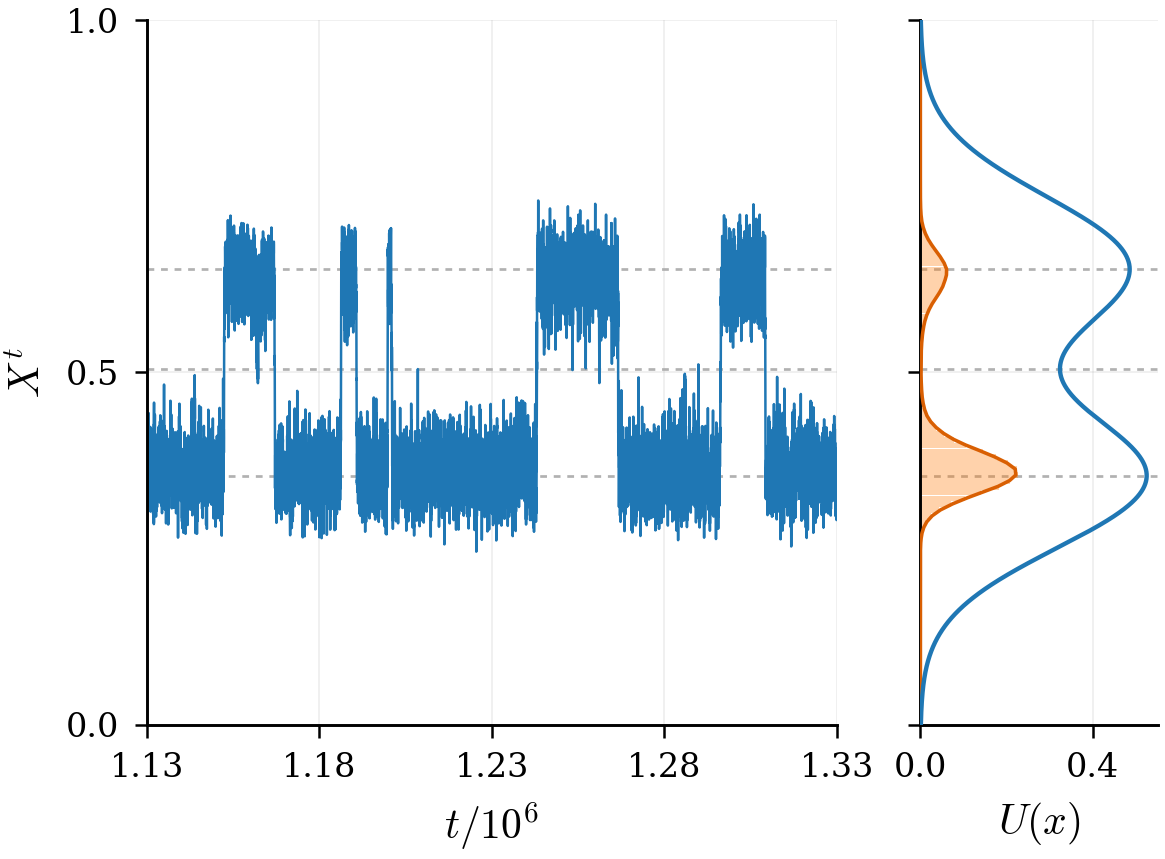}
  \caption{(Left) Mean opinion $X^t$ versus $t$ in a simulation with asymmetrically bimodal $U(x)$ [Eq.~(\ref{eq:asym_bimod})], $N=15$,  $\Delta=0.02$ and learning rate $\mu=0.1$. A single cluster forms and its mean opinion fluctuates around the two local maxima of $U(x)$. (Right) Utility function $U(x)$ (blue solid line) and empirical histogram for $X^t$ (orange shaded area).}
  \label{fig:truth_timeseries}
\end{figure}

The left panel of Fig.~\ref{fig:truth_timeseries} shows the mean opinion $X^t$ as a function of $t$ for an example with $N=15$, $\varepsilon=0.2$, $\mu=0.1$, $\Delta=0.02$, and a utility function given by
\begin{equation}
    U(x)=0.52U_1(x)+0.48U_2(x),
\label{eq:asym_bimod}
\end{equation}
where $U_1(x)$ and $U_2(x)$ are Gaussians with means at $x_L = 0.35$ and $x_R = 0.65$, respectively, and standard deviation $0.1$. The utility function $U(x)$ (blue solid line) and an empirical histogram for $X^t$ (orange shaded area) are shown in the right panel of Fig.~\ref{fig:truth_timeseries}. This example illustrates a case where the utility function has local maxima at two different opinion values. Although not shown in the figure, all agents form a single opinion cluster, i.e., $|x_i^t - x_j^t|<\varepsilon$ for all pairs $i$, $j$. This cluster appears to drift stochastically, spending the majority of time near the global maximum of $U(x)$ near $x=0.35$, even though the local maximum near $x=0.65$ has almost the same utility. In what follows, we will study the stochastic dynamics of a single cluster of opinions and will consider multiple clusters at the end of the paper.

\afterpage{
\begin{figure*}[t]
  \centering
  \includegraphics[width=\textwidth]{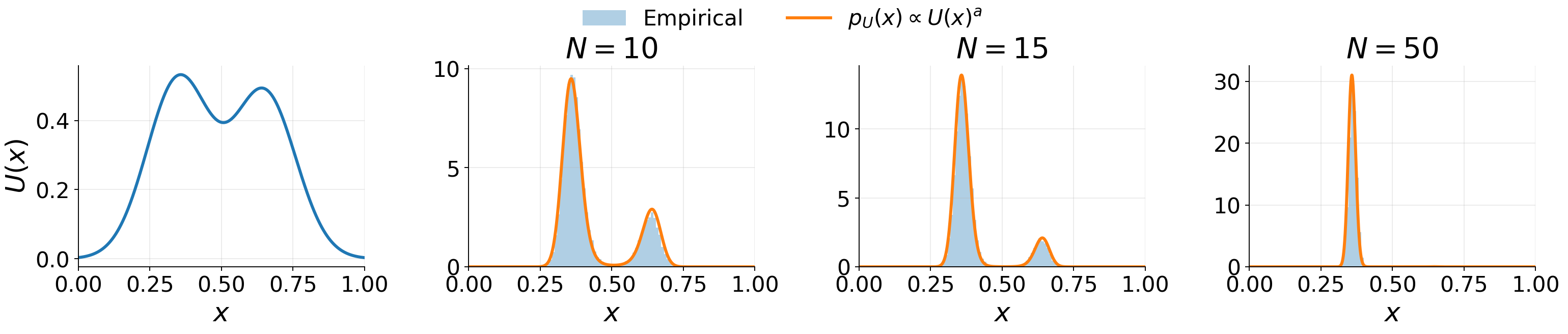}
  \caption{
  (Left panel) Bimodal utility function $U(x)$ [see Eq.~(\ref{eq:asym_bimod})].
  (Right three panels) Empirical steady-state histograms (light blue shaded areas) and theoretical steady-state curves of the population mean $X$ (orange solid curves) for $N=10,15,50$ at fixed noise standard deviation $\Delta=0.01$ and learning rate $\mu=1/2$. The theoretical stationary density $p^\ast(X)\propto U(X)^a$ is obtained from Eq.~\eqref{eq:stationary_main}.
  As $N$ increases, the theory predicts and the simulations confirm localization of the density around the global utility maximum.}
  \label{fig:Ua_overlay}
\end{figure*}
}

\paragraph{Single-cluster regime.}
We focus first on regimes in which the population forms a single cluster of opinions whose diameter remains below $\varepsilon$, so that an interaction occurs at every time step. Our goal is to derive a low-dimensional stochastic evolution equation for the mean opinion, defined as
\begin{equation}
X^t \equiv \frac{1}{N}\sum_{i=1}^N x_i^t.
\label{eq:mean}
\end{equation}
A small-spread expansion in $\delta x_i^t=x^t_i-X^t$ using $\Delta t= 1/N$ yields (see Supplementary Material) an effective drift for the mean,
\begin{equation}
A(x) \equiv \E\left[\frac{X^{t+1}- X^t}{\Delta t}\middle| X^t=x\right] 
\;\approx\;
2\mu\,\sigma^2\frac{U'(x)}{U(x)}\,,
\label{eq:drift_main}
\end{equation}
where $\sigma^2=\E[\delta x_i^2]$ is the cluster variance, and, to leading order, a noise-dominated diffusion coefficient
\begin{equation}
D_{\text{eff}}(x)
\equiv \frac{1}{2}\E\!\left[\frac{(X^{t+1}- X^t)^2}{\Delta t}\middle| X^t=x\right]
\;\approx\;
\frac{\Delta^2}{N}.
\label{eq:diff_main}
\end{equation}
The drift is proportional to $\frac{d}{dx}\ln U(x)$, and therefore the mean opinion moves, on average, in the direction that increases the utility function. The mean opinion satisfies the stochastic differential equation
\begin{equation}
dX_t
=
2\mu \sigma^2 \frac{U'(X_t)}{U(X_t)}\,dt
+
\Delta \sqrt{\frac{2}{N}}\, dW_t,
\label{eq:single_cluster_sde}
\end{equation}
where $W_t$ is a Wiener process.

Assuming that the evolution of the density $p(x,t)$ can be approximated by a Fokker--Planck equation, we write
\begin{equation}
\partial_t p
=
-\partial_x\!\big(A(x)p\big)+\partial_x^2\!\big(D_{\mathrm{eff}}p\big).
\label{eq:fp_mean}
\end{equation}
Using $A(x)$ and $D_\text{eff}$ from Eqs. \eqref{eq:drift_main} and \eqref{eq:diff_main}, the stationary solution $p^*(x)$ of \eqref{eq:fp_mean} vanishing at $x\rightarrow \pm \infty$, has the closed form
\begin{equation}
p^\ast(x)\propto U(x)^{a},
\qquad
a\equiv \frac{2\mu N\sigma^2}{\Delta^2}.
\label{eq:stationary_main}
\end{equation}
The constant $a$ depends on the cluster variance $\sigma^2$. To close the analysis, we find the cluster variance by considering the evolution of a representative agent's opinion $x^t_i$. Its drift and diffusion are given, to first order in $\delta x^t_i$, by
\begin{equation}
A_i(x) \equiv \E\left[\frac{X_i^{t+1}- X_i^t}{\Delta t}\middle| X^t=x\right] 
= -\,2\mu\,\delta x_i^t,
\label{eq:single_drift}
\end{equation}

\begin{equation}
D_{i}(x)
\equiv \frac{1}{2}\E\!\left[\frac{(X_i^{t+1}- X_i^t)^2}{\Delta t}\middle| X^t=x\right]
\approx2\mu^2\,\sigma^2 + \Delta^2\
,
\label{eq:single_diff}
\end{equation}

Solving the Fokker--Planck equation for the stationary density of agent $i$'s opinion, $p^*_i(x)$, we find that $p^*_i(x)$ is a Gaussian with mean $X^t$ and variance
\begin{equation}
\E[\delta x^2_i]
\equiv \sigma^2 = {\mu\sigma^2} + \frac{\Delta^2}{2\mu}\
.
\label{eq:single_var}
\end{equation}

Solving for $\sigma^2$, we find that
\begin{equation}
\sigma^2 = \frac{\Delta^2}{2\mu(1-\mu)}
.
\label{eq:single_dev}
\end{equation}
Inserting this in Eq. \eqref{eq:stationary_main}, we get

\begin{equation}
a=\frac{N}{1-\mu}.
\label{eq:a_main}
\end{equation}

One of our main results is, therefore, that the stationary distribution for the mean opinion is given by
\begin{equation}
p^*(x) \propto U(x)^{\frac{N}{1-\mu}},
\label{eq:reduced_steady_state}
\end{equation}
which is a Gibbs distribution for the potential $V(x) = - \ln(U(x))$ with inverse temperature $N/(1-\mu)$. Thus, the steady-state distribution for the mean opinion inherits the shape of the utility: regions where $U$ is large become more probable, with sharpness controlled by $a$. Remarkably, the stationary distribution $p^*(x)$ is independent of the properties of the microscopic noise term, including its variance $\Delta^2$. However, the variance of a typical agent's opinion is proportional to $\Delta^2$.

To illustrate this result, we consider again the utility function $U(x)$ in Eq.~\eqref{eq:asym_bimod}. This utility function has two local maxima at $x_L=0.35$ and $x_R=0.65$, with the value of $U(x)$ at $x=0.35$ being slightly larger than that at $x=0.65$ (see left panel of Fig.~\ref{fig:Ua_overlay}). Correspondingly, the stationary distribution of the mean opinion has local maxima at $x=0.35$ and $x=0.65$. However, as $N$ increases or $\mu$ approaches 1, the exponent $a$ increases and the stationary distribution $p^*(x)\propto U(x)^a$ localizes at the \textit{global} maximum at $x=0.35$. In the left panel of Fig.~\ref{fig:Ua_overlay} we show the utility function $U(x)$ as a function of $x$, and in the three remaining panels we show the theoretical expression for the stationary distribution (solid orange lines) and a histogram of the mean opinion values obtained by iterating Eq. \eqref{eq:update} numerically for 200,000 time steps with $\mu=1/2$ and Gaussian noise $\eta_i^t$ with $\Delta=0.01$. As expected, even moderately large values of $N$ result in localization of the stationary distribution around the local maximum.

The result expressed in Eq. \eqref{eq:reduced_steady_state} provides an explicit relation between the microscopic learning rate $\mu$, system size $N$, and the sharpness of the stationary distribution for the mean opinion. In particular, the exponent scales linearly with $N$, reflecting that larger populations produce tighter macroscopic fluctuations of the mean.

\paragraph{Utility Well Switching.}
As an illustration of our results, we will study the case where the utility function is bimodal, such as in Eq.~\eqref{eq:asym_bimod}. In this case, the stationary density \eqref{eq:stationary_main} is also bimodal, and the mean opinion can exhibit metastability - that is, it remains for long times near one local utility maximum and only occasionally crosses the intervening lower-utility region to the other local maximum. As noted previously, this behavior is visible in Fig.~\ref{fig:truth_timeseries}, along with the corresponding $U(x)$ and the steady-state distribution for the mean opinion. A natural question arises: what is the \textit{mean first-passage time} (MFPT) required for the mean opinion $X^t$ to escape one well?

Using the standard one-dimensional escape theory with the drift and diffusion in Eqs.~\eqref{eq:drift_main} and \eqref{eq:diff_main} gives the Arrhenius--Kramers estimate for the MFPT from the well near $x_L=0.35$ to a small neighborhood of the barrier near $x_M=0.5$ \cite{gardiner2009stochastic},
\begin{equation}
\log T_{x_L\to x_M}
\approx
\log C
+
\frac{N}{1-\mu}
\log\frac{U(x_L)}{U(x_M)},
\label{eq:mfpt_scaling}
\end{equation}
where $C$ is a constant prefactor depending on the curvature of the potential.
Equation \eqref{eq:mfpt_scaling} indicates that escape times grow exponentially in $N$ and, at fixed $N$, also increase rapidly as the learning rate $\mu$ increases.
The empirical dependence of the MFPT on $1/(1-\mu)$ is approximately linear on a logarithmic scale (see Supplementary Material), which is consistent with \eqref{eq:mfpt_scaling}. Thus the utility function not only determines the stationary distribution of the mean, but also controls the time scale of switching events between competing opinion wells.

\paragraph{Multiple Clusters.}
Above, we derived an effective one-dimensional stochastic description for the evolution of the mean opinion of a single cluster. Now we illustrate how this analysis can be used to efficiently simulate and, in some cases, analyze theoretically the evolution of multiple clusters. As a simple example, we consider the evolution of two clusters of agents of sizes $N/2$ each, such that the opinion of all agents in cluster 1 is initially 0 and the opinion of all agents in cluster 2 is initially 1. Furthermore, we assume that the utility function is a Gaussian centered at $\bar{x}=1/2$ with standard deviation $\beta$.
Initially, if the confidence bound $\varepsilon$ is small enough, the two clusters will evolve independently following the theory developed above for a single cluster. However, once the clusters get sufficiently close, they will interact. We assume that clusters quickly merge when the distance between their mean opinions $X^t_1$ and $X^t_2$ first reaches $\varepsilon$. Now we derive an expression for the expected time $T_m$ until the two clusters merge. For two clusters of sizes $N/2$ each and the Gaussian utility function $U(x)$ above, their mean opinions $X^t_1$, $X^t_2$ satisfy the stochastic differential equations
\begin{equation}
dX^t_1=\frac{1}{2} \frac{\Delta^2}{(1-\mu)\beta^2}(\bar{x}-X_1)dt+\Delta \sqrt{\frac{2}{N}} dW^1_t,
\label{eq:cluster_drift1}
\end{equation}

\begin{figure}[t]
  \centering
  \includegraphics[width=\columnwidth]{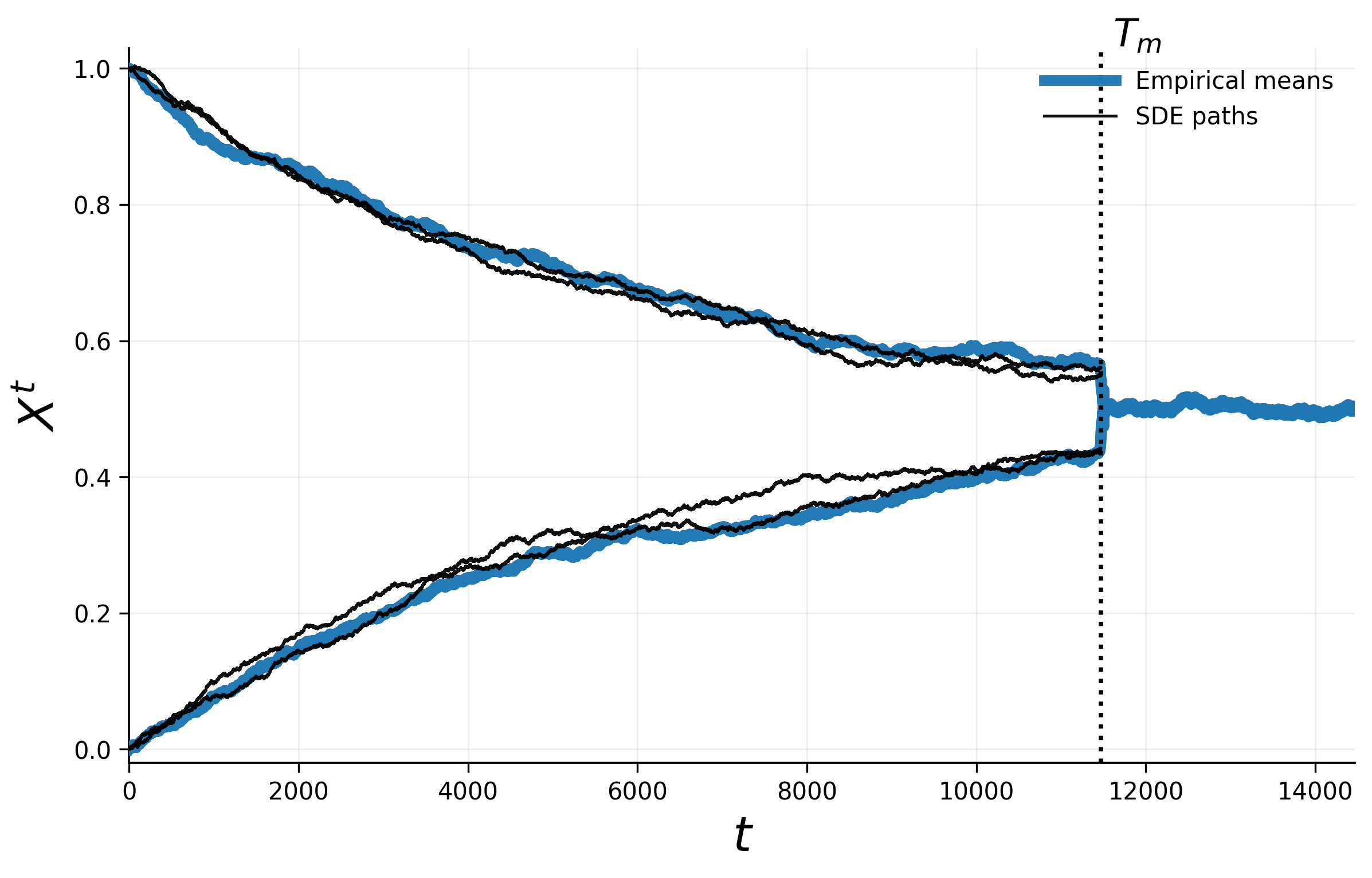}
  \caption{Comparison of simulated cluster mean trajectories (blue) and two simulated SDE paths (black).}
  \label{fig:cluster_merge}
\end{figure}

\begin{equation}
dX^t_2=\frac{1}{2} \frac{\Delta^2}{(1-\mu)\beta^2}(\bar{x}-X_2)dt+\Delta \sqrt{\frac{2}{N}} dW^2_t,
\label{eq:cluster_drift2}
\end{equation}
where $W^1_t$ and $W^2_t$ are independent Wiener processes. The factor of $1/2$ in the drift accounts for the reduced probability that two agents in the same cluster are selected for an update (a factor of $1/4$) and for the larger effect of a single agent on the cluster mean opinion (a factor of $2$), as shown in the Supplementary Material.
According to our approximate criterion for cluster merging, the merging time corresponds to the time when the Ornstein-Uhlenbeck variable $Z_t=X^t_2-X^t_1$  first reaches $\varepsilon$. Figure~\ref{fig:cluster_merge} shows the mean opinion of the two separate clusters, $X_1^t$ and $X_2^t$, calculated from direct numerical simulations of Eq.~(\ref{eq:update}) with $N = 50$, $\varepsilon = 0.1$, $\mu = 0.96$, $\Delta = 0.002$ (light blue curve), and two realizations of the stochastic differential equations (\ref{eq:cluster_drift1})-(\ref{eq:cluster_drift2}) (black curves). The figure illustrates that the stochastic differential equations describe well the evolution of each single cluster, and that the clusters do merge when the separation between their means is approximately $\varepsilon$.

\begin{figure}
  \centering
  \includegraphics[width=\columnwidth]{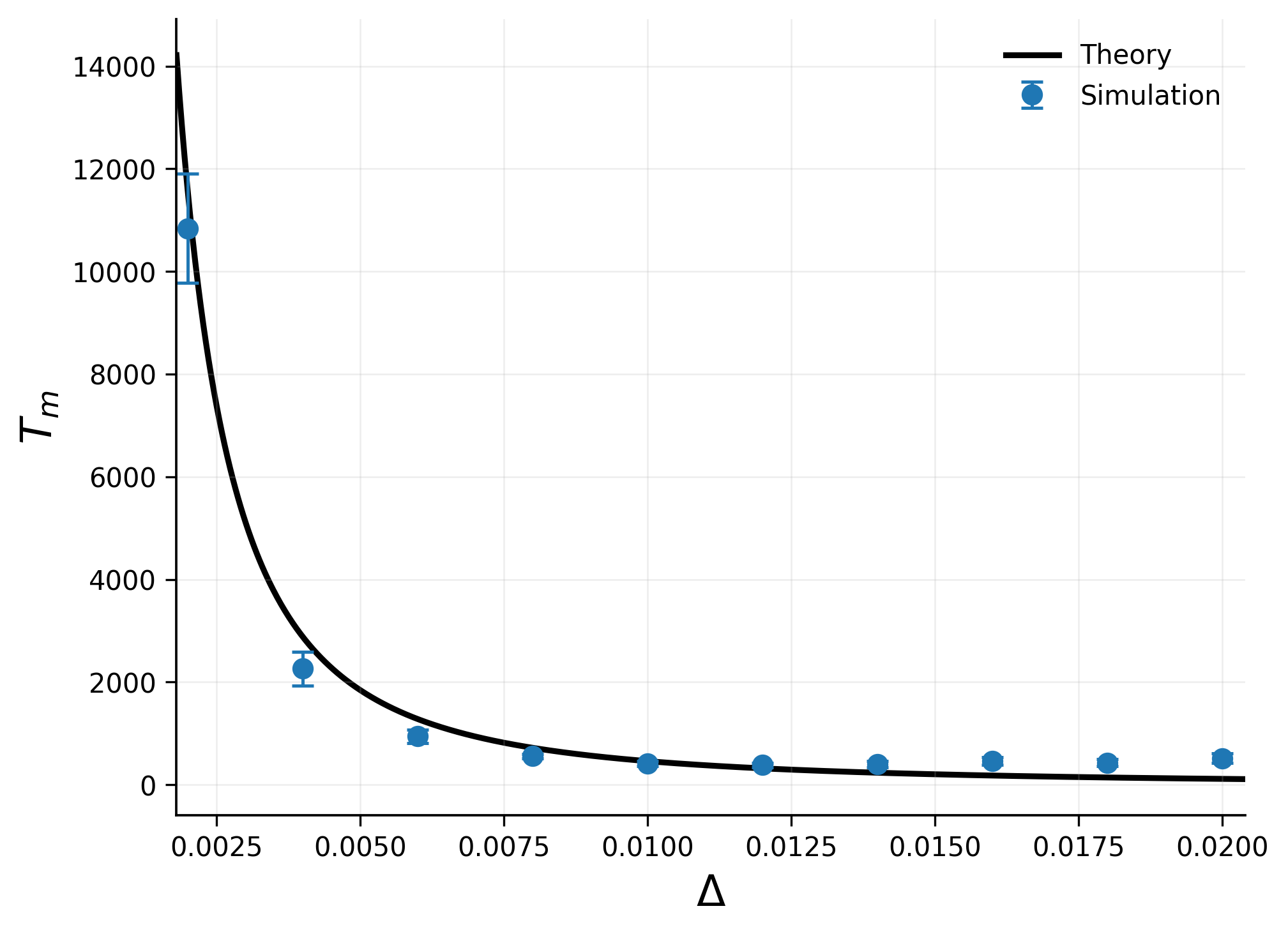}
  \caption{Numerically estimated average merging time $T_m$ (blue circles) and the theoretical value from Eq.~(\ref{eq:merge_time}) (solid line). Error bars indicate the empirical standard deviation. Agreement is strongest in the weak-noise regime where the assumptions leading to Eq. (\ref{eq:merge_time}) apply.}
  \label{fig:merge_time_comparison}
\end{figure}

A standard mean first-passage time calculation for weak noise gives (see the Supplementary Material for the derivation and a more general expression)
\begin{equation}
T_m=\frac{2(1-\mu)\beta^2}{\Delta^2}\ln\Big({\frac{1}{\varepsilon}\Big)}.
\label{eq:merge_time}
\end{equation}
In Fig.~\ref{fig:merge_time_comparison} we plot the numerically estimated average merging time $T_m$ (blue circles) as a function of $\Delta$, obtained by averaging over $30$ realizations, and the theoretical value from Eq.~(\ref{eq:merge_time}).

\paragraph{Discussion and outlook.}

In this Letter we introduced and studied a minimal model of opinion dynamics guided by the utility that agents derive from holding different opinions. We found that the evolution of a single opinion cluster follows a low-dimensional stochastic differential equation, with a drift toward higher values of the utility function. The utility function $U(x)$ determines the stationary distribution of the mean opinion, which is a Gibbs distribution with potential $-\ln(U(x))$ and inverse temperature $N/(1-\mu)$.  For multiple clusters,  bounded confidence creates an interaction bottleneck: clusters initially evolve independently under utility-driven drift until their separation reaches the confidence threshold, after which rapid merger of the clusters occurs. This separation of time scales enables a reduced stochastic description of cluster evolution and leads to analytical estimates for merging and escape times.

We restricted our attention to utility functions that depend only on the current value of the agent's opinion, i.e., $U_i({\bf x}^t) = U(x_i^t)$. However, one can also consider  a situation where agents derive social utility from holding an opinion that conforms to other agents' opinions, such as those of  their neighbors in a social network. More generally, an important feature of our framework is that various social or epistemic situations can be modeled using a utility function that rewards both choosing an objectively better opinion 
%[such as the value $x = 0.35$ for the utility function in  Eq.~(\ref{eq:asym_bimod})] 
and conforming to others' opinions. This framework would be a natural  model for phenomena such as scientific communities with competing theories, financial or political belief formation under social reinforcement, institutional consensus formation, public-health belief dynamics, cult-like or high-conformity group dynamics, and other settings in which beliefs are shaped both by evidence and by social payoff.

\begin{acknowledgments}
We thank Erin Obermayer, Poom Kritpracha, and Chanin Kumpeerakij for helpful discussions.
\end{acknowledgments}

\bibliographystyle{apsrev4-2}
\bibliography{sample}

% ================================================================
% Supplementary Material
% ================================================================

\clearpage
\onecolumngrid
\appendix

\setcounter{equation}{0}
\renewcommand{\theequation}{S\arabic{equation}}

\section*{Supplementary Material}

Here, we give the derivations for the expressions presented in the main text. We assume that the population remains in a single cluster whose diameter is smaller than the confidence bound, and use the update rule
\begin{equation}
x^{t+1}_{i}
=
x^t_i
+2\mu \frac{U_j^t}{U_i^t+U_j^t}(x^t_j-x^t_i)
+\eta^t_i ,
\label{eq:S_update}
\end{equation}
and analogously for $j$, with
\begin{equation}
\E[\eta_i^t]=0,
\qquad
\E[\eta_i^t\eta_j^{t'}]=\Delta^2\delta_{ij}\delta_{tt'},
\label{eq:S_noise}
\end{equation}
where $\delta_{ij}$ is the Kronecker delta.

\section{Mean-opinion drift}

Define the mean opinion
\begin{equation}
X^t\equiv\frac{1}{N}\sum_{i=1}^N x_i^t.
\label{eq:S_mean_def}
\end{equation}
Suppose that at time $t$ the chosen pair is $(i,j)$. Since only these two agents change their opinions,
\begin{equation}
X^{t+1}-X^t
=
\frac{1}{N}\Big[(x_i^{t+1}-x_i^t)+(x_j^{t+1}-x_j^t)\Big].
\label{eq:S_mean_increment_start}
\end{equation}
Using Eq.~\eqref{eq:S_update},
\begin{align}
x_i^{t+1}-x_i^t
&=
2\mu \frac{U_j^t}{U_i^t+U_j^t}(x_j^t-x_i^t)+\eta_i^t,
\label{eq:S_n_increment}
\\
x_j^{t+1}-x_j^t
&=
2\mu \frac{U_i^t}{U_i^t+U_j^t}(x_i^t-x_j^t)+\eta_j^t.
\label{eq:S_m_increment}
\end{align}
Hence
\begin{align}
X^{t+1}-X^t
&=
\frac{2\mu}{N}
\frac{U_j^t(x_j^t-x_i^t)-U_i^t(x_i^t-x_j^t)}{U_i^t+U_j^t}
+\frac{\eta_i^t+\eta_j^t}{N}
\notag\\
&=
\frac{2\mu}{N}
\frac{U_j^t-U_i^t}{U_i^t+U_j^t}(x_j^t-x_i^t)
+\frac{\eta_i^t+\eta_j^t}{N}.
\label{eq:S_mean_increment_exact}
\end{align}
Therefore, since the noise has zero mean and recalling $\Delta t = 1/N$, the conditional drift is
\begin{equation}
A(x)\equiv \E\left[\frac{X^{t+1}-X^t}{\Delta t}\middle| X^t=x\right]
=
2\mu\,
\E\!\left[
\frac{U_j^t-U_i^t}{U_i^t+U_j^t}(x_j^t-x_i^t)
\,\middle|\, X^t=x
\right].
\label{eq:S_drift_exact}
\end{equation}
Now write the single-cluster decomposition
\begin{equation}
x_i^t=X^t+\delta x_i^t,
\qquad
\frac{1}{N}\sum_{i=1}^N \delta x_i^t =0,
\qquad
\sigma^2:=\E[(\delta x_i^t)^2].
\label{eq:S_cluster_decomp}
\end{equation}
Expanding the utility to first order around the mean,
\begin{equation}
U(X^t+\delta x_n^t)
=
U(X^t)+U'(X^t)\,\delta x_n^t+O((\delta x_n^t)^2).
\label{eq:S_U_expand}
\end{equation}
Using this for $n=i,j$ gives
\begin{align}
\frac{U_j^t-U_i^t}{U_i^t+U_j^t}
&\approx
\frac{U'(X^t)(\delta x_j^t-\delta x_i^t)}
{2U(X^t)+U'(X^t)(\delta x_i^t+\delta x_j^t)}
\notag\\
&\approx
\frac{U'(X^t)}{2U(X^t)}(\delta x_j^t-\delta x_i^t),
\label{eq:S_ratio_expand}
\end{align}
to first order in the cluster spread. Since
\begin{equation}
x_j^t-x_i^t=\delta x_j^t-\delta x_i^t,
\label{eq:S_diff_delta}
\end{equation}
Eq.~\eqref{eq:S_drift_exact} becomes
\begin{align}
A(x)
&\approx
2\mu\,
\frac{U'(x)}{2U(x)}
\E\!\left[(\delta x_j^t-\delta x_i^t)^2 \,\middle|\, X^t=x\right],
\label{eq:S_drift_prefinal}
\end{align}
where we used $\E[\eta_i^t+\eta^t_j]=0$.
Assuming that $\delta x^t_i, \delta x^t_j$ are uncorrelated (strictly, they are weakly anti-correlated because their sum vanishes, but this introduces only finite N corrections), we have $\E[\delta x^t_i \delta x^t_j]=0$, so
\begin{equation}
\E\!\left[(\delta x_j^t-\delta x_i^t)^2\right]
=
\E[(\delta x_j^t)^2]+\E[(\delta x_i^t)^2]-2\E[\delta x_j^t\delta x_i^t]
\approx 2\sigma^2,
\label{eq:S_cluster_variance_identity}
\end{equation}
so that
\begin{equation}
A(x)
\approx
2\mu\,\frac{U'(x)}{U(x)}\,\sigma^2.
\label{eq:S_mean_drift_final}
\end{equation}
This is the result given in Eq.~(5) of the main text.

\section{Mean-opinion diffusion}

Define the effective diffusion coefficient
\begin{equation}
D_{\mathrm{eff}}(x)
:=
\frac{1}{2}\E\!\left[\frac{(X^{t+1}-X^t)^2}{\Delta t}\,\middle|\,X^t=x\right].
\label{eq:S_Deff_def}
\end{equation}
From Eq.~\eqref{eq:S_mean_increment_exact},
\begin{align}
D_{\mathrm{eff}}(x)
&=
\frac{N}{2}
\E\!\left[
\left(
\frac{2\mu}{N}
\frac{U_j^t-U_i^t}{U_i^t+U_j^t}(x_j^t-x_i^t)
+
\frac{\eta_i^t+\eta_j^t}{N}
\right)^2
\middle|\,X^t=x
\right].
\label{eq:S_Deff_expand_start}
\end{align}
Because the noise has zero mean and is independent of the exact configuration at time $t$, the cross terms vanish after conditioning on $X^t=x$. Thus
\begin{align}
D_{\mathrm{eff}}(x)
&=
\frac{2\mu^2}{N}
\E\!\left[
\left(
\frac{U_j^t-U_i^t}{U_i^t+U_j^t}
\right)^2
(x_j^t-x_i^t)^2
\middle|\,X^t=x
\right]
+
\frac{1}{2N}
\E\!\left[(\eta_i^t+\eta_j^t)^2\right].
\label{eq:S_Deff_split}
\end{align}
The deterministic term is higher order in the cluster spread. Indeed, using Eq.~\eqref{eq:S_ratio_expand},
\begin{equation}
\left(
\frac{U_j^t-U_i^t}{U_i^t+U_j^t}
\right)^2
(x_j^t-x_i^t)^2
=
O\!\left((\delta x)^4\right),
\label{eq:S_deterministic_quartic}
\end{equation}
so to leading order it is negligible compared with the additive noise term. Since
\begin{equation}
\E[(\eta_i^t+\eta_j^t)^2]
=
\E[(\eta_i^t)^2]+\E[(\eta_j^t)^2]
=
2\Delta^2,
\label{eq:S_noise_sum}
\end{equation}
we obtain
\begin{equation}
D_{\mathrm{eff}}(x)
\approx
\frac{\Delta^2}{N}.
\label{eq:S_Deff_final}
\end{equation}
This is the diffusion coefficient quoted in Eq.~(6) of the main text.

\section{Stationary density of the mean opinion}

We approximate the slow evolution of the density $p(x,t)$ of the mean opinion by using a Fokker--Planck equation,
\begin{equation}
\partial_t p
=
-\partial_x\!\big(A(x)p\big)
+\partial_x^2\!\big(D_{\mathrm{eff}}(x)p\big),
\label{eq:S_FP_mean}
\end{equation}
and using Eqs.~\eqref{eq:S_mean_drift_final} and \eqref{eq:S_Deff_final}, we have
\begin{equation}
\partial_t p
=
-\partial_x\!\left(
2\mu\frac{U'(x)}{U(x)}\sigma^2\,p
\right)
+
\partial_x^2\!\left(
\frac{\Delta^2}{N}p
\right).
\label{eq:S_FP_mean_substituted}
\end{equation}
At stationarity, imposing zero probability flux gives
\begin{equation}
2\mu\frac{U'(x)}{U(x)}\sigma^2\,p^\ast(x)
=
\frac{d}{dx}\left(\frac{\Delta^2}{N}p^\ast(x)\right).
\label{eq:S_zero_flux}
\end{equation}
This reduces to
\begin{equation}
\frac{{p^{\ast}}'(x)}{p^\ast(x)}
=
\frac{2\mu N\sigma^2}{\Delta^2}\,\frac{U'(x)}{U(x)}.
\label{eq:S_log_derivative}
\end{equation}
Integrating,
\begin{align}
\ln p_\ast(x)
&=
\frac{2\mu N\sigma^2}{\Delta^2}\ln U(x)+\mathrm{const},
\label{eq:S_logp_integrated}
\\
p_\ast(x)
&\propto
U(x)^a,
\qquad
a\equiv\frac{2\mu N\sigma^2}{\Delta^2}.
\label{eq:S_stationary_mean_final}
\end{align}
This is the stationary law given in Eq.~(9) of the main text.

\section{Representative-agent drift}

We now derive the self-consistency relation for $\sigma^2$ by studying a representative agent. Fix an agent $i$. The probability that agent $i$ is selected in a randomly chosen pair is
\begin{equation}
\mathbb{P}(i\ \text{is chosen})
=
\frac{N-1}{\binom{N}{2}}
=
\frac{2}{N}.
\label{eq:S_choose_prob}
\end{equation}
Therefore
\begin{align}
\E[x_i^{t+1}-x_i^t]
&=
\frac{2}{N}\,
\E\!\left[x_i^{t+1}-x_i^t\,\middle|\,i\ \text{chosen}\right].
\label{eq:S_agent_drift_condition}
\end{align}
Conditioned on agent $i$ being chosen and paired with agent $j$,
\begin{equation}
x_i^{t+1}-x_i^t
=
2\mu\frac{U_j^t}{U_i^t+U_j^t}(x_j^t-x_i^t)+\eta_i^t.
\label{eq:S_agent_increment_exact}
\end{equation}
Using the same first-order expansion as before,
\begin{equation}
\frac{U_j^t}{U_i^t+U_j^t}
\approx
\frac{1}{2},
\label{eq:S_ratio_half}
\end{equation}
so
\begin{align}
\E\!\left[x_i^{t+1}-x_i^t\,\middle|\,i\ \text{chosen}\right]
&\approx
\mu\,\E[\delta x_j^t-\delta x_i^t]
\notag\\
&=
-\mu\,\delta x_i^t,
\label{eq:S_agent_conditional_drift}
\end{align}
because $\E[\delta x_j^t\mid i \   \text{chosen}]=0$ by the symmetry of the cluster. Hence,
\begin{equation}
\E\left[\frac{x_i^{t+1}-x_i^t}{\Delta t}\right]
=
-2\mu\,\delta x_i^t.
\label{eq:S_agent_drift_final}
\end{equation}
This is the single-agent drift quoted in Eq.~(10) of the main text.

\section{Representative-agent diffusion}

Define the single-agent diffusion coefficient by
\begin{equation}
D_i(x)
:=
\frac{1}{2}\E\!\left[\frac{(x_i^{t+1}-x_i^t)^2}{\Delta t}\,\middle|\,X^t=x\right].
\label{eq:S_Di_def}
\end{equation}
Conditioning on whether agent $i$ is chosen,
\begin{align}
D_i(x)
&=
\frac{1}{2}\frac{2}{N}\frac{1}{\Delta t}\E\!\left[
\left(
2\mu\frac{U_j^t}{U_i^t+U_j^t}(x_j^t-x_i^t)+\eta_i^t
\right)^2
\middle|\,X^t=x,\ i\ \text{chosen}
\right].
\label{eq:S_Di_conditioned}
\end{align}
Using again $U_j^t/(U_i^t+U_j^t)\approx 1/2$ and simplifying the prefactor,
\begin{align}
D_i(x)
&\approx
\E\!\left[
\left(
\mu(\delta x_j^t-\delta x_i^t)+\eta_i^t
\right)^2
\middle|\,X^t=x
\right]=
\mu^2\E[(\delta x_j^t-\delta x_i^t)^2]
+
\E[(\eta_i^t)^2],
\label{eq:S_Di_expand}
\end{align}
where the cross term vanishes because $\E[\eta_i^t]=0$. Using Eq.~\eqref{eq:S_cluster_variance_identity},
\begin{equation}
\E[(\delta x_j^t-\delta x_i^t)^2]\approx 2\sigma^2,
\label{eq:S_delta_diff_sq}
\end{equation}
and Eq.~\eqref{eq:S_noise}, and then recalling that $\E[(\eta_i^t)^2]=\Delta^2$, 
\begin{equation}
D_i(x)
\approx
2\mu^2\sigma^2+\Delta^2.
\label{eq:S_Di_final}
\end{equation}
This is the single-agent diffusion formula corresponding to Eq.~(11) in the main text.

\section{Single-agent stationary density and variance closure}

To leading order, the representative agent experiences a linear drift toward the cluster center with approximately constant diffusion,
\begin{equation}
A_i(x)=-2\mu\,\delta x,
\qquad
D_i=2\mu^2\sigma^2+\Delta^2.
\label{eq:S_OU_coeffs}
\end{equation}
The corresponding Fokker--Planck equation for the density $p_i(\delta x,t)$ is
\begin{equation}
\partial_t p_i
=
-\partial_{\delta x}\!\big(A_i\,p_i\big)
+
\partial_{\delta x}^2\!\big(D_i\,p_i\big).
\label{eq:S_FP_agent}
\end{equation}
At stationarity,
\begin{equation}
-2\mu\,\delta x\,p_i(\delta x)
=
\frac{d}{d\delta x}\!\left[
\left(2\mu^2\sigma^2+\Delta^2
\right)p_i(\delta x)
\right].
\label{eq:S_agent_zero_flux}
\end{equation}
Since $D_i$ is constant, this becomes
\begin{equation}
\frac{p_i'(\delta x)}{p_i(\delta x)}
=
-\frac{2\mu\,\delta x}{2\mu^2\sigma^2+\Delta^2}.
\label{eq:S_agent_log_derivative}
\end{equation}
Integrating results in a Gaussian,
\begin{align}
p_i(\delta x)
&\propto
\exp\!\left(
-\frac{\mu\,\delta x^2}{2\mu^2\sigma^2+\Delta^2}
\right)
\notag\\
&=
\exp\!\left(
-\frac{\delta x^2}{2\left[\mu\sigma^2+\frac{\Delta^2}{2\mu}\right]}
\right).
\label{eq:S_agent_gaussian}
\end{align}
Matching this with the standard Gaussian form
\begin{equation}
p_i(\delta x)\propto \exp\!\left(-\frac{\delta x^2}{2\sigma^2}\right),
\label{eq:S_gaussian_match}
\end{equation}
gives the self-consistency condition
\begin{equation}
\sigma^2
=
\mu\sigma^2+\frac{\Delta^2}{2\mu}.
\label{eq:S_variance_self_consistency}
\end{equation}
Solving for $\sigma^2$,
\begin{equation}
\sigma^2
=
\frac{\Delta^2}{2\mu(1-\mu)}.
\label{eq:S_sigma_final}
\end{equation}
Substituting Eq.~\eqref{eq:S_sigma_final} into Eq.~\eqref{eq:S_stationary_mean_final} gives
\begin{equation}
a
=
\frac{2\mu N}{\Delta^2}\cdot \frac{\Delta^2}{2\mu(1-\mu)}
=
\frac{N}{1-\mu}.
\label{eq:S_a_final}
\end{equation}
Therefore the stationary density of the mean opinion is
\begin{equation}
p_\ast(x)\propto U(x)^{\frac{N}{1-\mu}},
\label{eq:S_reduced_stationary}
\end{equation}
which is the result quoted in Eq.~(15) of the main text.

\section{Mean first-passage time for bimodal utility functions}

Here we assume that the utility function is bimodal, with two local maxima at $x = x_L, x_R$ separated by an intervening minimum at $x = x_M$. Eq.~(\ref{eq:single_cluster_sde}) from the main text can be rewritten as
\begin{equation}
dX_t = - V'(X_t)\,dt +\Delta \sqrt{\frac{2}{N}}\, dW_t,
\end{equation}
where
\begin{align}
V(x) = -2\mu \sigma^2 \ln U(x).
\end{align}
The potential $V(x)$ has local minima at $x =x_L,x_R$ and an intervening local maximum at $x = x_M$. 
Standard Kramers escape theory gives the asymptotic scaling for the mean first-passage time from the local minimum $x = x_L$ to $x = x_M$  
\begin{equation}
T_{x_L\to x_M}
\sim
C \exp\left(
\frac{V(x_M)-V(x_L)}{D}
\right),
\end{equation}
where $C$ is a prefactor depending on the local curvature of the potential near $x_M$ and $x_L$ and $D=\Delta^2/N$ is the effective diffusion coefficient of the mean opinion. Using the expression for $V(x)$,
\begin{equation}
V(x_M)-V(x_L)
=
2\mu\sigma^2
\ln\left(
\frac{U(x_L)}{U(x_M)}
\right).
\end{equation}
Substituting this into the Kramers formula and using
\begin{equation}
\sigma^2
=
\frac{\Delta^2}{2\mu(1-\mu)},
\end{equation}
we obtain
\begin{equation}
\log T_{x_L\to x_M}
\approx
\log C
+
\frac{N}{1-\mu}
\log\left(
\frac{U(x_L)}{U(x_M)}
\right),
\end{equation}
which is Eq.~(\ref{eq:mfpt_scaling}).

\section{Drift and diffusion for separate clusters}

We consider two clusters of sizes $N/2$,
$C_1=\{1,2,\ldots,N/2\}$
and
$C_2=\{N/2+1,N/2+2,\ldots,N\}$
with initial opinions
$x_i^0=1$ if $i\in C_1$
and
$x_i^0=0$ if $i\in C_2$.
We define the mean cluster opinions
\[
X_1^t=\frac{1}{(N/2)}\sum_{i\in C_1}x_i^t,
\qquad
X_2^t=\frac{1}{(N/2)}\sum_{i\in C_2}x_i^t .
\]

Proceeding as before, we calculate the drift and diffusion for the mean
cluster opinions, but now we need to condition on the selection of two agents
from the cluster
\[
A_1(x)=
\mathbb{E}\left[
\left.
\frac{X_1^{t+1}-X_1^t}{\Delta t}
\,\right|\, X_1^t=x
\right]
\]
\[
=
\mathbb{E}\left[
\left.
\frac{X_1^{t+1}-X_1^t}{\Delta t}
\,\right|\,
X_1^t=x,\; i,j \text{ selected from cluster }1
\right]
\times
\mathbb{P}(i,j \text{ selected from cluster }1)
\]
\[
=
\frac{1}{4}
\mathbb{E}\left[
\left.
\frac{X_1^{t+1}-X_1^t}{\Delta t}
\,\right|\,
X_1^t=x,\; i,j \text{ selected from cluster }1
\right] .
\]
Using that
\[
X_1^{t+1}-X_1^t
=
\frac{2}{N}
\left[
(x_i^{t+1}-x_i^t)
+
(x_j^{t+1}-x_j^t)
\right]
\]
and proceeding as before, we recover the drift in Eqs.~(17) and (18) of the
main text.

For the effective diffusion, we have
\[
D_1(x)
\equiv
\frac{1}{2}
\mathbb{E}\left[
\left.
\frac{(X_1^{t+1}-X_1^t)^2}{\Delta t}
\,\right|\, X_1^t=x
\right]
\]
\[
=
\frac{1}{2}
\mathbb{E}\left[
\left.
\frac{(X_1^{t+1}-X_1^t)^2}{\Delta t}
\,\right|\,
X_1^t=x,\; i,j \text{ selected from cluster }1
\right]
\times
\mathbb{P}(i,j \text{ selected from cluster }1)
\]
\[
=
\frac{1}{2}\times \frac{1}{4}\times
\mathbb{E}\left[
\left.
\frac{(X_1^{t+1}-X_1^t)^2}{\Delta t}
\,\right|\,
X_1^t=x,\; i,j \text{ selected from cluster }1
\right] .
\]
Again, using that
\[
X_1^{t+1}-X_1^t
=
\frac{2}{N}
\left[
(x_i^{t+1}-x_i^t)
+
(x_j^{t+1}-x_j^t)
\right],
\]
we obtain
\[
D_1(x)=\frac{\Delta^2}{N}=D_2(x),
\]
as used in Eqs.~(17) and (18) of the main text.

\section{Mean merging time}

Defining
\[
Z_t=X_t^2-X_t^1,
\]
where $X_t^1$ and $X_t^2$ satisfy Eqs.~(17) and (18), $Z_t$ satisfies the
Ornstein--Uhlenbeck process
\[
dZ_t
=
-\frac{1}{2}\frac{\Delta^2}{(1-\mu)\beta^2}Z_t dt
+
\Delta\sqrt{\frac{4}{N}}\,dW_t,
\]
where $W_t$ is a Wiener process. The initial condition is $Z_0=1$, and the
(approximate) merging time $T_m$ is the first time when $Z_t=\varepsilon$.
In the deterministic $(N\to\infty)$ limit, one can find the merging time
$T_m$ by solving
\[
Z_{T_m}
=
e^{-\frac{1}{2}\frac{\Delta^2}{(1-\mu)\beta^2}T_m}
=
\varepsilon,
\]
which gives the expression in Eq.~(\ref{eq:merge_time}) in the main text.

More generally, the mean first-passage time $T(z)$ from $z$ to
$\varepsilon$ satisfies the Backward Kolmogorov equation
\[
-1
=
-\frac{1}{2}\frac{\Delta^2}{(1-\mu)\beta^2}z T'(z)
+
\frac{2\Delta^2}{N}T''(z),
\qquad
T(\varepsilon)=0 .
\]
For $z=1$, the mean first-passage time $\mathbb{E}[T_m]=T(1)$ is given by
\[
\mathbb{E}[T_m]
=
\frac{N}{2\Delta^2}
\frac{\sqrt{\pi}}{2}
\,
\beta
\sqrt{\frac{8(1-\mu)}{N}}
\int_{\varepsilon}^{1}
e^{\frac{N z^2}{8(1-\mu)\beta^2}}
\operatorname{erfc}
\left(
\sqrt{\frac{N}{8(1-\mu)\beta^2}}\,z
\right)
\,dz .
\]

\section{Scaling of Mean First-Passage Time}

\begin{figure}
  \centering
  \includegraphics[width=0.75\columnwidth]{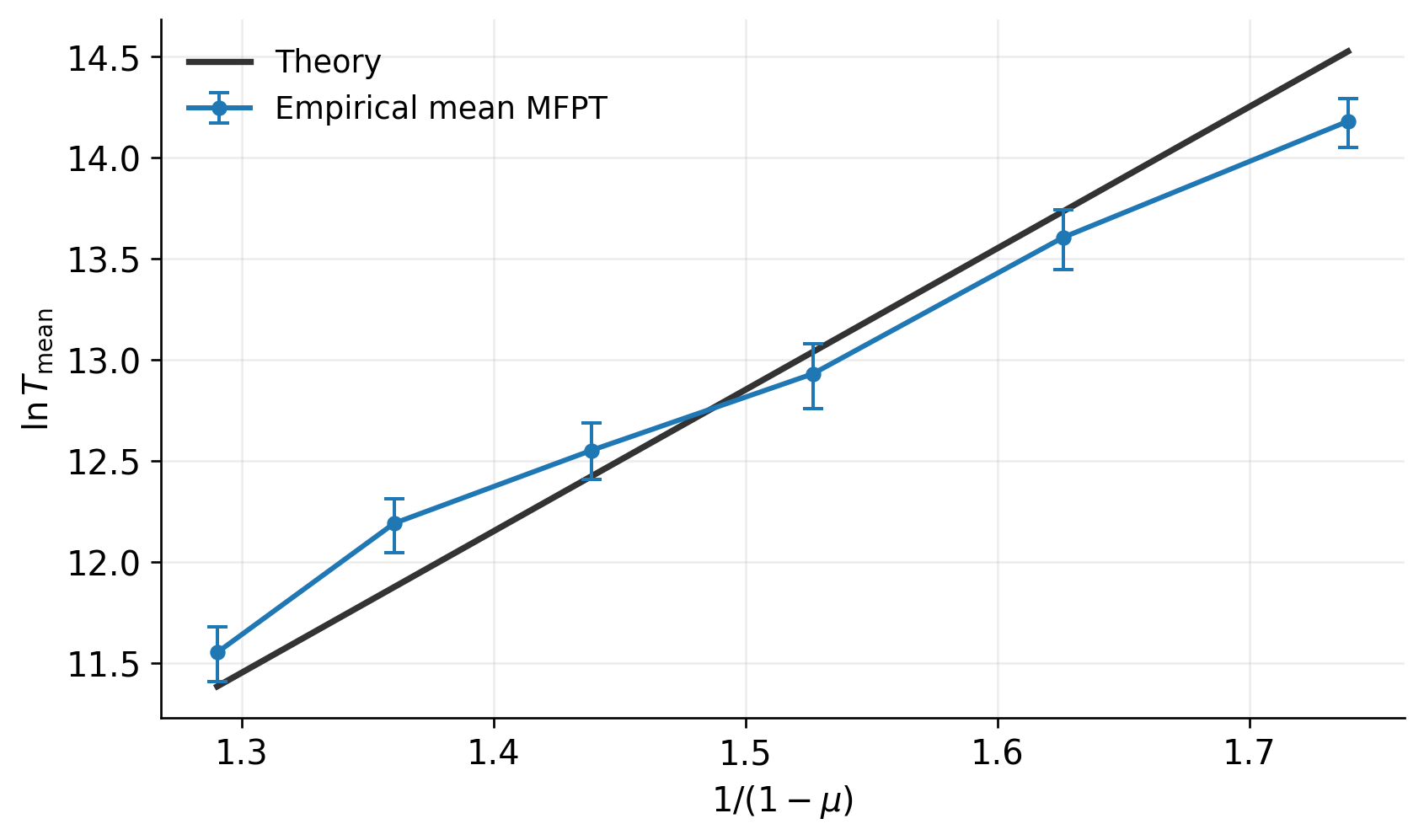}
  \caption{Comparison of theoretical and empirical $\mu$ dependence with error bars corresponding to the empirical standard deviation. Approximate linearity in $1/(1-\mu)$ indicates alignment between theory and simulation.}
  \label{fig:mu_dependence}
\end{figure}

In the main text we found that the MFPT from one local maximum of the utility function to another scales like
\begin{equation}
\log T_{x_L\to x_M}
\approx
\log C
+
\frac{N}{1-\mu}
\log\frac{U(x_L)}{U(x_M)}.
\label{eq:mfpt_scaling_SM}
\end{equation}
In order to test this scaling, we consider a symmetric, bimodal utility function $U(x)$. For a fixed $\mu$, we start a simulation of the model in Eq.~(\ref{eq:update}) with all the agents' opinions in one maximum of $U(x)$, and run it until the mean opinion $X^t$ reaches the midpoint between the two maxima. We record this time as $T_{x_L\to x_M}$ and repeat the process, averaging $40$ realizations to obtain $T_{\text{mean}}$. Figure~\ref{fig:mu_dependence} shows the empirical  (symbols) and theoretical (solid line) $\ln(T_{\text{mean}})$ versus $1/(1-\mu)$. 

\end{document}